\documentclass{article}

\usepackage{microtype}
\usepackage{graphicx}
\usepackage{subcaption}
\usepackage{booktabs} 
\usepackage{multirow}
\usepackage{caption}
\usepackage{wrapfig,lipsum}
\usepackage{arydshln}
\usepackage[usenames,dvipsnames]{xcolor}
\usepackage{textcomp}
\usepackage{scalerel}
\usepackage{bbding}
\usepackage[most]{tcolorbox}
\usepackage{fancyvrb}
\usepackage{fvextra}
\usepackage{multicol}

\usepackage{hyperref}
\usepackage{pifont}

\usepackage{amsmath,amsfonts,bm}

\usepackage{xspace}
\usepackage{xcolor}
\usepackage{graphicx}
\usepackage{nicematrix}

\newcommand\Comment[1]{}

\newcommand{\ours}{SWE-Replay\xspace}
\newcommand{\gemini}{\scalebox{1}{\scalerel*{\includegraphics{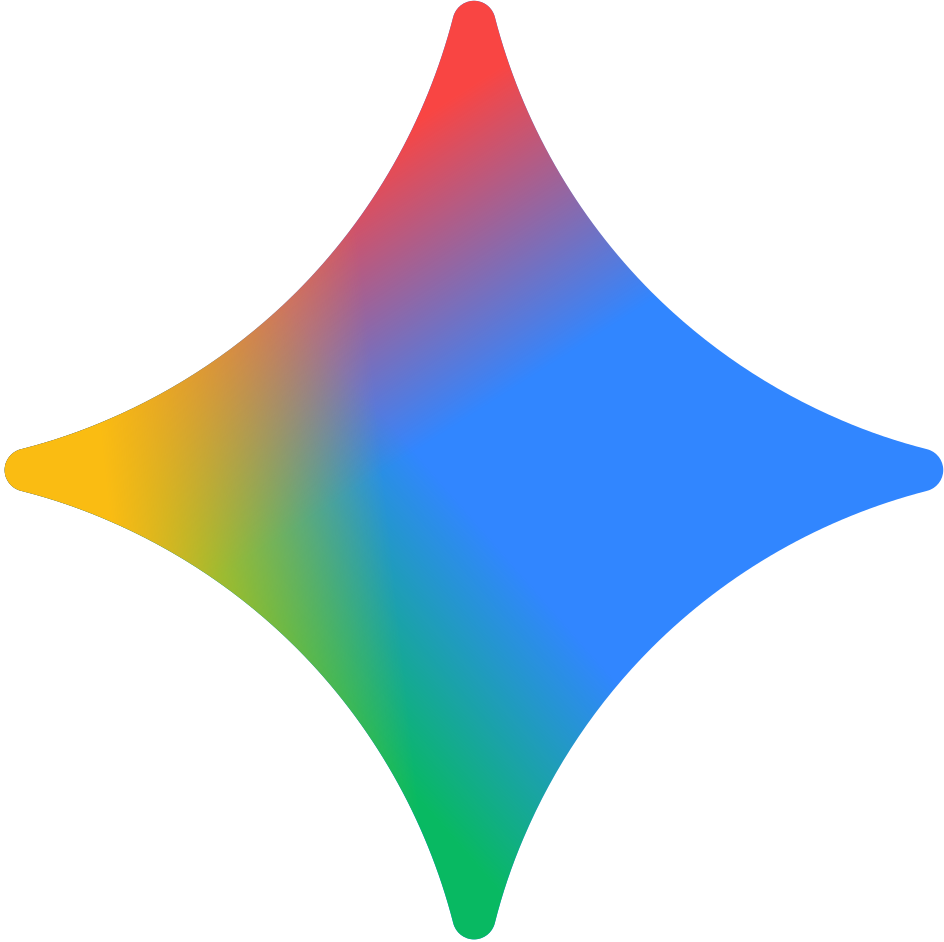}}{\textrm{C}}}\xspace}
\newcommand{\mistral}{\scalebox{1}{\scalerel*{\includegraphics{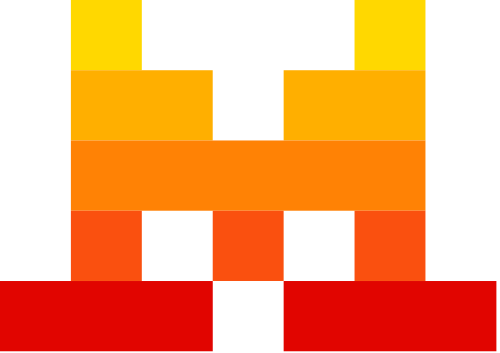}}{\textrm{C}}}\xspace}
\newcommand{\xmark}{\ding{55}}

\newtcolorbox{promptbox}[1]{
  enhanced,
  breakable,
  boxrule = 1.5pt,
  fontupper = \normalsize,
  fonttitle = \large\bf\color{white},
  arc = 5pt,
  rounded corners,
  colframe=blue!50!green,
  colback=blue!5!white,
  coltitle=white,
  title = #1,
  left=4pt %
}

\newcommand{\eg}{\emph{e.g.,}\xspace}
\newcommand{\ie}{\emph{i.e.,}\xspace}

\def\eqref#1{equation~\ref{#1}}

\def\1{\bm{1}}

\DeclareMathAlphabet{\mathsfit}{\encodingdefault}{\sfdefault}{m}{sl}
\SetMathAlphabet{\mathsfit}{bold}{\encodingdefault}{\sfdefault}{bx}{n}

\usepackage[preprint]{icml2026}

\usepackage{amsmath}
\usepackage{amssymb}
\usepackage{mathtools}
\usepackage{amsthm}

\usepackage{enumitem}

\usepackage[capitalize,noabbrev]{cleveref}

\theoremstyle{plain}

\theoremstyle{definition}

\theoremstyle{remark}

\usepackage[textsize=tiny]{todonotes}

\icmltitlerunning{\ours: Efficient Test-Time Scaling for Software Engineering Agents}

\begin{document}

\twocolumn[
  \icmltitle{\ours: Efficient Test-Time Scaling for Software Engineering Agents}



  \icmlsetsymbol{equal}{*}

  \begin{icmlauthorlist}
    \icmlauthor{Yifeng Ding}{uiuc}
    \icmlauthor{Lingming Zhang}{uiuc}
  \end{icmlauthorlist}

  \icmlaffiliation{uiuc}{Siebel School of Computing and Data Science, University of Illinois Urbana-Champaign, USA}

  \icmlcorrespondingauthor{Yifeng Ding}{yifeng6@illinois.edu}

  \icmlkeywords{Machine Learning, ICML}

  \vskip 0.3in
]



\printAffiliationsAndNotice{}  

\begin{abstract}
Test-time scaling has been widely adopted to enhance the capabilities of Large Language Model (LLM) agents in software engineering (SWE) tasks. However, the standard approach of repeatedly sampling trajectories from scratch is computationally expensive. While recent methods have attempted to mitigate costs using specialized value agents, they can suffer from model miscalibration and fail to generalize to modern agents that synthesize custom bash scripts as tools. In this paper, we introduce \ours, the first efficient and generalizable test-time scaling technique for modern agents without reliance on potentially noisy value estimates. \ours optimizes the scaling process by recycling trajectories from prior trials, dynamically choosing to either explore from scratch or exploit archived experience by branching at critical intermediate steps. This selection of intermediate steps is driven by the potential and reasoning significance of repository exploration, rather than external LLM-based quality estimates. Our evaluation shows that, on SWE-Bench Verified, \ours consistently outperforms naive scaling, reducing costs by up to 17.4\% while maintaining or even improving performance by up to 3.8\%. Further evaluation on SWE-Bench Pro and Multilingual validates the generalizability of \ours, establishing it as a robust foundation for efficient test-time scaling of software engineering agents.
\end{abstract}

\section{Introduction}\label{sec:introduction}
Large Language Models (LLMs) have evolved significantly, progressing from basic code completion tools~\cite{austin2021program,chen2021codex,li2022competition,ding-etal-2024-mathcal,ding-etal-2025-planning} to sophisticated interactive agents capable of navigating repositories, executing tests, and submitting patches end-to-end~\cite{yang2024swe,wang2024openhands,zhang2024autocoderover,liu2024large,xia2024agentless,carbonneaux2025cwm}. Modern agentic frameworks, such as SWE-agent~\cite{yang2024swe} and OpenHands~\cite{wang2024openhands}, equip LLMs with tools like terminals, editors, and search engines to tackle complex repositories. To further enhance capabilities in software engineering (SWE) tasks, a primary driver of progress has been \textit{test-time scaling}~\cite{zhang2025survey}, which increases inference-time computation to yield higher-quality solutions. Existing works have demonstrated that generating multiple candidate solutions with a positive temperature and selecting a final answer via test-based feedback improves performance log-linearly with the number of samples, establishing a promising scaling law for SWE tasks~\cite{brown2024large,ehrlich2025codemonkeys}.

\begin{figure*}[t]
\centering
\includegraphics[width=0.78\linewidth]{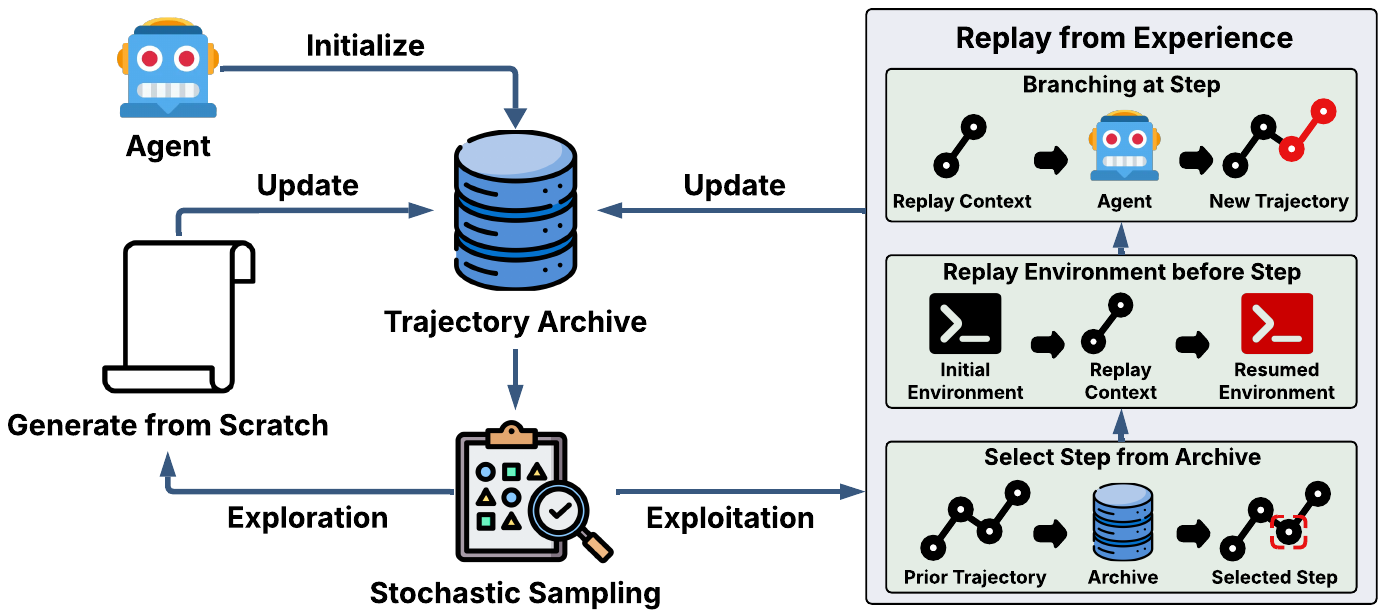}
\caption{Overview of \ours.}
\label{fig:overview}
\vspace*{-1.5\baselineskip}
\end{figure*}

Despite this promise, test-time scaling is computationally expensive due to the cost of repeated sampling from scratch~\cite{kim2026costdynamicreasoningdemystifying}. Consequently, several existing works have been proposed to mitigate these costs. SWE-Search~\cite{antoniades2024swe} proposed using a modified Monte Carlo Tree Search (MCTS) to prune unpromising trajectories early, relying on a value agent to estimate the quality of each step. Similarly, Satori-SWE~\cite{zeng2025satorisweevolutionarytesttimescaling} proposes to achieve sample-efficient test-time scaling by prompting agents to self-improve the scores of their prior generations estimated by a reward model. However, these approaches have the following limitations: quality scores from value agents and reward models can be compromised by model miscalibration~\cite{son2024llm}, introducing noise during scaling; more importantly, existing approaches are tailored to pipeline-based scaffolds (\eg Moatless\footnote{https://github.com/aorwall/moatless-tools}) and cannot generalize to modern agentic frameworks like SWE-agent~\cite{yang2024swe}. Specifically, existing works design specific prompts to evaluate pre-defined tools in the pipeline, such as a structured search/retrieval tool. However, modern agents are designed to synthesize custom bash scripts without templates. This flexibility makes it infeasible to design tool-specific evaluation prompts, rendering existing works incompatible with modern agents.

To bridge this gap, we introduce \ours, the first \textit{efficient} and \textit{generalizable} test-time scaling technique designed to reduce trajectory sampling costs while improving trajectory quality for modern agents, \textit{without reliance on LLM-as-a-Judge}. Our core insight is to recycle previously sampled trajectories by resuming exploration at carefully selected intermediate steps, rather than generating every trajectory from scratch. As shown in Figure~\ref{fig:overview}, \ours maintains an archive of sampled trajectories and performs stochastic sampling iteratively: either explore by sampling a new trajectory from scratch or exploit by replaying from the middle of an existing trajectory. In the exploitation phase, \ours identifies a critical step $s_t$ based on its potential to explore new search space and its reasoning intensity, efficiently restores its environment state, and branches at step $s_t$ by sampling a new step $s'_t$ to replace $s_t$ while continuing exploration. This mechanism bypasses reliance on potentially inaccurate LLM-as-a-Judge, generalizes naturally to modern agentic scaffolds, and employs a streamlined select-and-replay mechanism to ensure scalability.

We evaluate \ours on the widely used SWE-Bench Verified\footnote{https://openai.com/index/introducing-swe-bench-verified/}, and the more complex SWE-Bench Pro~\cite{deng2025swe} and Multilingual\footnote{https://www.swebench.com/multilingual.html}. Our results show that, on SWE-Bench Verified, \ours consistently reduces the cost of naive test-time scaling by up to 17.4\% while maintaining or even improving performance by up to 3.8\%, across three different LLM backends and two different agentic scaffolds. On SWE-Bench Pro and Multilingual, \ours further demonstrates its consistent generalizability to diverse SWE problems. In addition, our analysis reveals that \ours successfully shifts exploration to the long-tail of repository files, as visualized in Figure \ref{fig:distribution}. Finally, we provide an interesting theoretical intuition on how replaying optimizes the quality of exploration in \ours.

In summary, we make the following contributions:

\vspace*{-0.8\baselineskip}

\begin{itemize}[leftmargin=1em]
    \setlength{\parskip}{0pt}
    \item We present \ours, the first efficient test-time scaling technique for software engineering agents that reduces trajectory sampling costs while improving trajectory quality, \textbf{without any reliance on LLM-as-a-Judge} and \textbf{generalizable to any modern agent scaffold}.
    \item On SWE-Bench Verified, \ours consistently reduces the sampling cost by up to 17.4\% while maintaining or even improving the resolve rate by up to 3.8\%, across different agent scaffolds and LLM backends. We further validate the generalization of \ours to diverse software issues on SWE-Bench Pro and Multilingual.
    \item Our in-depth analysis provides both \textit{empirical} results showing that \ours enables agents to explore more diverse repository spaces, and \textit{theoretical} intuition that bridges the gap between the performance gain and the effectiveness of step selection in \ours.
\end{itemize}

\begin{figure}[t]
\centering
\includegraphics[width=0.9\linewidth]{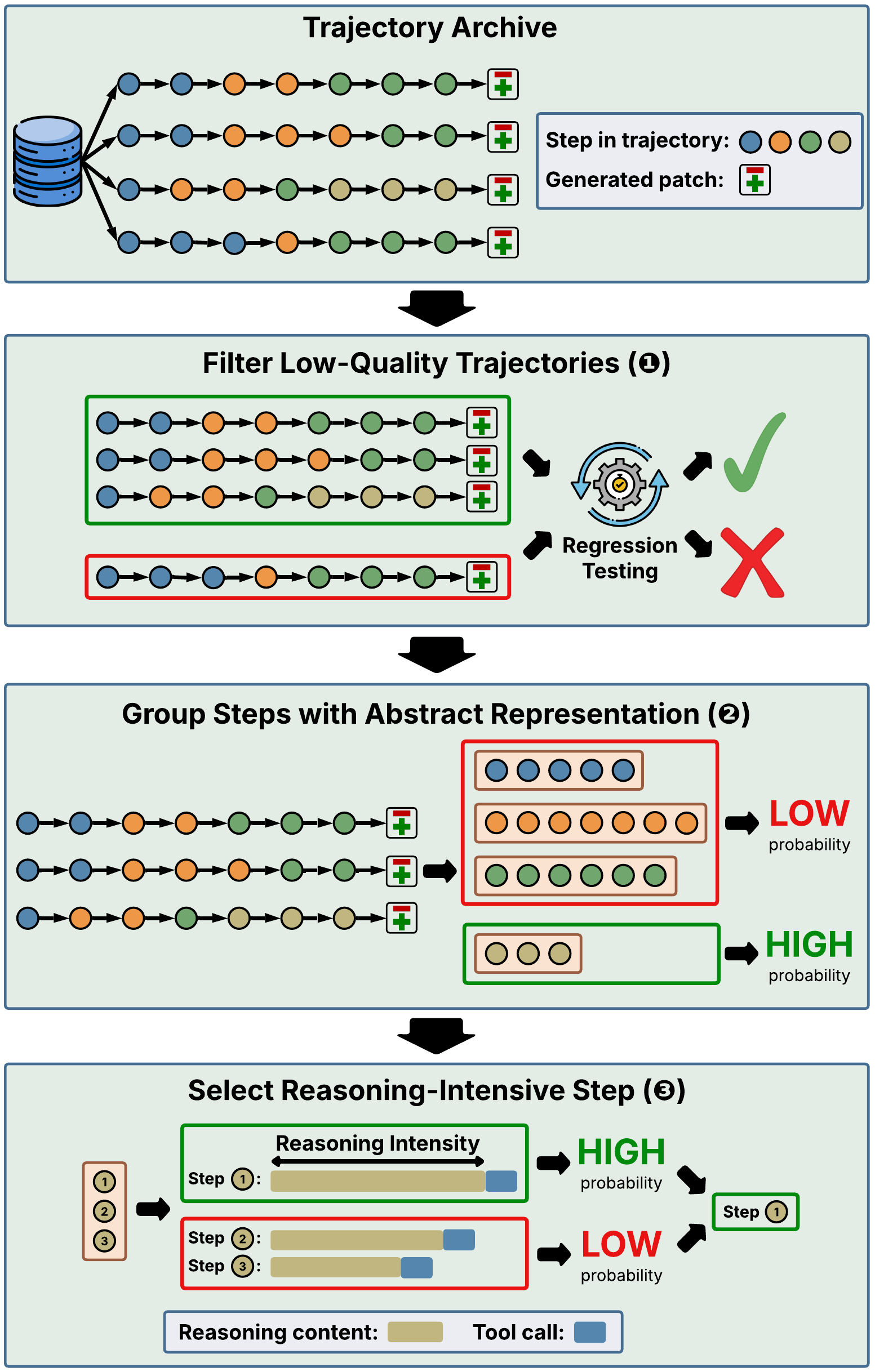}
\caption{Overview of step selection in \ours.}
\label{fig:step_selection}
\vspace*{-\baselineskip}
\end{figure}

\section{\ours}\label{sec:method}
To enable efficient and effective exploration for software engineering agents, we propose \ours, a novel algorithm that reduces the cost of trajectory sampling while improving trajectory quality. The key idea of \ours is to reuse previously sampled trajectories by resuming exploration at carefully selected intermediate steps, rather than repeatedly generating trajectories from scratch.

As shown in Figure \ref{fig:overview}, \ours maintains an archive of previously sampled trajectories and updates the archive with newly generated ones. Specifically, \ours initializes the archive with a single trajectory generated from scratch. It then iteratively conducts a stochastic sampling to decide whether to explore by generating a new trajectory from scratch or to exploit existing trajectories by resuming from intermediate steps in the archive. To exploit existing trajectories, \ours will select a critical step $s_t$ from the archive based on its potential to explore new repository space and its reasoning importance (\S\ref{sec:select}), restore the environment state before this step (\ie $s_{1,2,\cdots,t-1}$) efficiently (\S\ref{sec:return}), and resume exploration by generating a new step $s'_t$ to replace $s_t$ for new branches (\S\ref{sec:explore}).

\begin{figure}[t]
\centering
\includegraphics[width=0.85\linewidth]{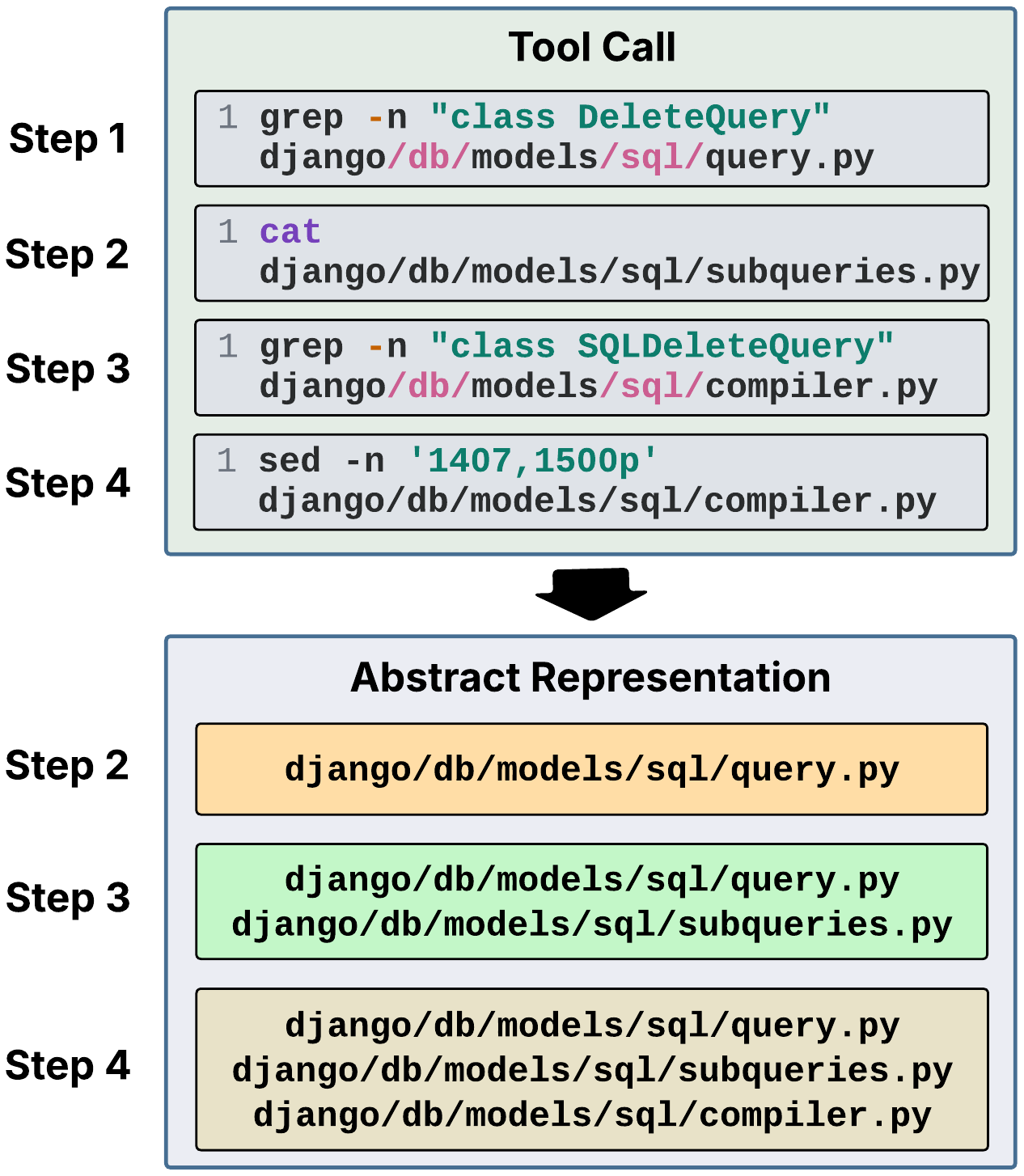}
\caption{Example of tool calls and their file-level representations. Step 1 is skipped as no file has been explored before it.}
\label{fig:step_representation_example}
\vspace*{-\baselineskip}
\end{figure}

\subsection{Selecting Steps}\label{sec:select}
Reliable identification of critical intermediate steps is central to efficient exploration. Intuitively, resuming at such steps allows the agent to revisit valuable regions of the search space that \textbf{are not explored comprehensively} and \textbf{require more reasoning efforts} from the agent.

To this end, \ours employs a hierarchical selection pipeline consisting of four components, as shown in Figure \ref{fig:step_selection}: (1) filtering low-quality trajectories from the archive (\S\ref{sec:select_filter}), (2) grouping steps using abstract state representations and selecting the critical group (\S\ref{sec:select_abstract}), (3) selecting concrete steps within a group based on reasoning intensity (\S\ref{sec:select_reasoning}), and (4) balancing exploration and exploitation through stochastic sampling (\S\ref{sec:select_adaptive}).

\subsubsection{Filtering Low-Quality Trajectories}\label{sec:select_filter}
To ensure that resumed exploration originates from reliable trajectories, \ours first removes low-quality trajectories from the archive. Trajectory quality is evaluated based on the final generated patch. In particular, we check whether the patch causes any existing regression tests to fail, and \textbf{discard all trajectories whose final patches introduce regression failures}. The motivation is that trajectories leading to regression failures often contain misleading intermediate steps, such as exploring irrelevant files, which can propagate the same errors when reused. In practice, we follow Agentless~\cite{xia2024agentless} to obtain regression tests for each issue. Since the original repositories do not provide reproduction tests and LLM-generated tests may be unreliable, \ours exclusively relies on existing regression tests from the repository, as shown in Figure \ref{fig:step_selection} (\ding{182}).

\subsubsection{Grouping Steps via Abstraction}\label{sec:select_abstract}
A key challenge in selecting critical steps is to distinguish meaningfully different exploration steps while avoiding fragmentation caused by superficial differences. Although one could represent each step using the raw environment output, such representations are typically long and nearly unique, making similarity assessment ineffective. Instead, \ours introduces a lightweight yet effective abstraction by \textbf{representing each step using the set of repository files explored before that step}, as shown in Figure \ref{fig:step_representation_example}.

We refer to this representation as a \textit{state}. Each state corresponds to a subset of steps that have explored the same files. Let $\{s_1, s_2, \dots, s_n\}$ denote the set of distinct abstract states extracted from all steps, where state $s_i$ corresponds to $v_i$ concrete steps. To prioritize states that access rarely explored regions of the repository, \ours assigns a sampling probability $p_i$ to each state based on its rarity:
\vspace*{-0.5\baselineskip}
\begin{equation}
p_i = \mathrm{softmax}\!\left(\frac{1}{v_i}\right)
= \frac{e^{\frac{1}{v_i}}}{\sum_{j=1}^{n} e^{\frac{1}{v_j}}}.
\label{eq:select_state}
\end{equation}
\vspace*{-1.2\baselineskip}

This formulation favors states reached by fewer steps, but still maintains a balanced trade-off between exploration and exploitation through sampling rather than greedily choosing the most rare state. Softmax normalization is adopted to maintain the sampling diversity for better performance with a flatter distribution, as validated in \S\ref{sec:ablation_normalization}.

\begin{figure}[t]
\centering
\includegraphics[width=0.9\linewidth]{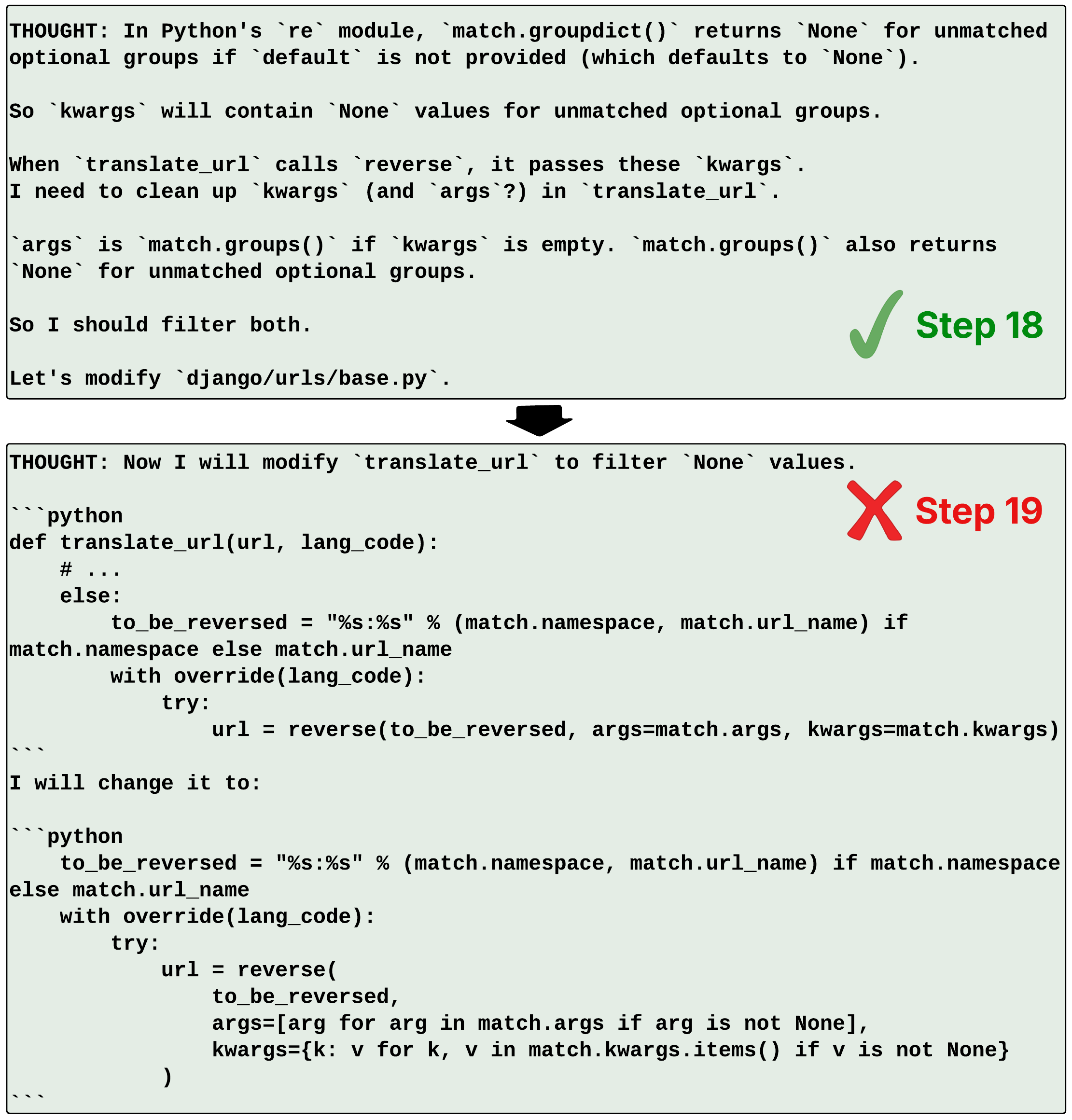}
\caption{Example of reasoning contents with different intensity.}
\label{fig:step_reasoning_example}
\vspace*{-1.5\baselineskip}
\end{figure}

\subsubsection{Detecting Reasoning-intensive Steps}\label{sec:select_reasoning}
After selecting a critical abstract state, \ours must choose a concrete step at which to resume exploration. Typically, modern agents are designed to perform explicit reasoning at each step. Inspired by this, we select the final step by \textbf{prioritizing reasoning-intensive ones}, as shown in Figure \ref{fig:step_selection} (\ding{184}). It is based on the intuition that steps requiring more deliberation are often \textit{critical decision points} in the task. Therefore, \ours branches at such steps to explore alternative solutions by sampling new steps to replace them.

To measure the reasoning intensity, one straightforward idea is to use the raw reasoning length. However, the raw length cannot effectively convey the reasoning intensity, as the reasoning content may include copies of raw code snippets or concrete inputs, which significantly extend the length without providing any meaningful reasoning gains. As shown in Figure \ref{fig:step_reasoning_example}, while step 19 has longer reasoning content, its reasoning consists of two detailed but less informative edit drafts; in contrast, step 18 contains multiple short, insightful paragraphs analyzing the program logic. Based on this observation, \ours instead uses the number of paragraphs in the reasoning content as a \textit{robust structural proxy} to measure the reasoning intensity. Empirical ablation (\S\ref{sec:ablation_reasoning}) confirms that paragraph count outperforms raw token length (60.0\% vs 58.0\%), validating that structural segmentation captures reasoning quality better than verbosity.

Given the $v_i$ steps $\{s_{i,1}, \dots, s_{i,v_i}\}$ within state $s_i$, with corresponding numbers of reasoning paragraphs $\{l_{i,1}, \dots, l_{i,v_i}\}$, similar to \S\ref{sec:select_abstract}, \ours assigns sampling probabilities as follows:
\vspace*{-0.9\baselineskip}
\begin{equation}
p_{i,j} = \mathrm{softmax}\!\left(l_{i,j}\right) = \frac{e^{l_{i,j}}}{\sum_{j=1}^{v_i} e^{l_{i,j}}}.
\label{eq:select_step}
\end{equation}

\subsubsection{Balancing Exploration and Exploitation}\label{sec:select_adaptive}
To support effective test-time scaling, \ours balances exploration (sampling trajectories from scratch) and exploitation (resuming from prior trajectories) as follows:

\vspace*{-0.8\baselineskip}
\begin{itemize}[leftmargin=1em]
    \setlength{\parskip}{2pt}
    \item \textbf{Initialization (Exploration):} The first run always generates a trajectory from scratch using only the original issue description, forming the initial archive.
    \item \textbf{Scaling (Exploration vs.\ Exploitation):} For subsequent runs, exploration and exploitation are treated as independent Bernoulli trials with probability $p=0.5$. The strategy for each run is then sampled stochastically.
\end{itemize}
\vspace*{-0.8\baselineskip}

This design enables \ours to naturally trade off coverage and reuse as scaling proceeds.

\begin{table*}[]
\caption{Results on SWE-Bench Verified.}
\vspace*{-0.3\baselineskip}
\centering
\resizebox{1.9\columnwidth}{!}{
\begin{tabular}{@{}cllcccc@{}}
\toprule
\multirow{2}{*}{Agent Scaffold} & \multicolumn{1}{c}{\multirow{2}{*}{LLM Backend}} & \multicolumn{1}{c}{\multirow{2}{*}{Scaling Method}} & \multicolumn{1}{c}{Performance}            & \multicolumn{3}{c}{Efficiency}                                                                                          \\ \cmidrule(l){4-7} 
                                & \multicolumn{1}{c}{}                       & \multicolumn{1}{c}{}                                & \multicolumn{1}{c}{\% Resolved $\uparrow$} & \multicolumn{1}{c}{\# Input (k) $\downarrow$} & \multicolumn{1}{c}{\# Output (k) $\downarrow$} & \multicolumn{1}{c}{Avg. Cost (\$) $\downarrow$} \\ \midrule
\multirow{9}{*}{mini-SWE-agent} & \multirow{3}{*}{\gemini{} Gemini-2.5-Pro}            & Naive Scaling                                       & 58.0\%                                     & 3464.5                                  & 98.3                                     & 1.52                               \\
                                &                                            & \ours                                & $\,$\textbf{60.2\%}                            & \textbf{2980.4}                         & \textbf{86.0}                            & \textbf{1.32}                      \\
                                \addlinespace[0.1em]
                                \cdashline{3-7}
                                \addlinespace[0.3em]
                                &                                            & Improvement                               & \textcolor{MidnightBlue}{\textbf{$\uparrow$ 3.8\%}}                            & \textcolor{ForestGreen}{\textbf{$\downarrow$ 14.0\%}}                         & \textcolor{ForestGreen}{\textbf{$\downarrow$ 12.5\%}}                           & \textcolor{ForestGreen}{\textbf{$\downarrow$ 13.2\%}}                     \\ \cmidrule(l){2-7} 
                                & \multirow{3}{*}{\mistral{} Devstral-Small-2}          & Naive Scaling                                       & 62.2\%                                     & 14914.9                                 & 130.0                                    & 1.53                               \\
                                &                                            & \ours                                & $\,$\textbf{63.2\%}                            & \textbf{13221.4}                        & \textbf{111.2}                           & \textbf{1.36}                      \\
                                \addlinespace[0.1em]
                                \cdashline{3-7}
                                \addlinespace[0.3em]
                                &                                            & Improvement                               & \textcolor{MidnightBlue}{\textbf{$\uparrow$ 1.6\%}}                            & \textcolor{ForestGreen}{\textbf{$\downarrow$ 11.4\%}}                          & \textcolor{ForestGreen}{\textbf{$\downarrow$ 14.5\%}}                           & \textcolor{ForestGreen}{\textbf{$\downarrow$ 11.1\%}}                     \\ \cmidrule(l){2-7} 
                                & \multirow{3}{*}{\gemini{} Gemini-3-Pro}              & Naive Scaling                                       & 75.4\%                                     & 6849.2                                  & 106.4                                    & 2.88                               \\
                                &                                            & \ours                                & $\,$\textbf{75.6\%}                            & \textbf{5874.3}                         & \textbf{85.4}                            & \textbf{2.38}                      \\
                                \addlinespace[0.1em]
                                \cdashline{3-7}
                                \addlinespace[0.3em]
                                &                                            & Improvement                               & \textcolor{MidnightBlue}{\textbf{$\uparrow$ 0.3\%}}                             & \textcolor{ForestGreen}{\textbf{$\downarrow$ 14.2\%}}                          & \textcolor{ForestGreen}{\textbf{$\downarrow$ 19.7\%}}                           & \textcolor{ForestGreen}{\textbf{$\downarrow$ 17.4\%}}                     \\ \midrule
\multirow{6}{*}{Live-SWE-agent} & \multirow{3}{*}{\mistral{} Devstral-Small-2}          & Naive Scaling                                       & 63.2\%                                     & 14993.9                                 & 134.9                                    & 1.54                               \\
                                &                                            & \ours                                & $\,$\textbf{65.0\%}                            & \textbf{13233.0}                        & \textbf{115.7}                           & \textbf{1.36}                      \\
                                \addlinespace[0.1em]
                                \cdashline{3-7}
                                \addlinespace[0.3em]
                                &                                            & Improvement                               & \textcolor{MidnightBlue}{\textbf{$\uparrow$ 2.8\%}}                             & \textcolor{ForestGreen}{\textbf{$\downarrow$ 11.7\%}}                          & \textcolor{ForestGreen}{\textbf{$\downarrow$ 14.2\%}}                           & \textcolor{ForestGreen}{\textbf{$\downarrow$ 11.7\%}}                     \\ \cmidrule(l){2-7} 
                                & \multirow{3}{*}{\gemini{} Gemini-3-Pro}              & Naive Scaling                                       & 75.6\%                                     & 6948.7                                  & 107.0                                    & 2.93                               \\
                                &                                            & \ours                                & $\,$\textbf{75.8\%}                            & \textbf{6395.0}                         & \textbf{91.4}                            & \textbf{2.59}                      \\
                                \addlinespace[0.1em]
                                \cdashline{3-7}
                                \addlinespace[0.3em]
                                &                                            & Improvement                               & \textcolor{MidnightBlue}{\textbf{$\uparrow$ 0.3\%}}                             & \textcolor{ForestGreen}{\textbf{$\downarrow$ 8.0\%}}                           & \textcolor{ForestGreen}{\textbf{$\downarrow$ 14.6\%}}                           & \textcolor{ForestGreen}{\textbf{$\downarrow$ 11.6\%}}                     \\ \bottomrule
\end{tabular}
}
\label{tab:main}
\vspace*{-\baselineskip}
\end{table*}

\begin{table}[]
\caption{Results on SWE-Bench Pro and SWE-Bench Multilingual.}
\vspace*{-0.3\baselineskip}
\centering
\resizebox{0.9\columnwidth}{!}{
\begin{tabular}{@{}clcc@{}}
\toprule
Benchmark                               & \multicolumn{1}{c}{Scaling Method} & \% Resolved $\uparrow$ & Avg. Cost (\$) $\downarrow$ \\ \midrule
\multirow{3}{*}{Pro}          & Naive Scaling                      &      28.73\%       &      1.29        \\
                                        & \ours               &     $\,$\textbf{28.97\%}        &       \textbf{1.25}        \\
                                        \addlinespace[0.1em]
                                        \cdashline{2-4}
                                        \addlinespace[0.3em]
                                        & Improvement                        &      \textcolor{MidnightBlue}{\textbf{$\uparrow$ 0.8\%}}       &      \textcolor{ForestGreen}{\textbf{$\downarrow$ 3.1\%}}          \\ \midrule
\multirow{3}{*}{Multilingual} & Naive Scaling                      &      31.0\%       &       1.22         \\
                                        & \ours               &    $\,$\textbf{38.0\%}         &       \textbf{1.11}         \\
                                        \addlinespace[0.1em]
                                        \cdashline{2-4}
                                        \addlinespace[0.3em]
                                        & Improvement                        &   \textcolor{MidnightBlue}{\textbf{$\uparrow$ 22.6\%}}          &  \textcolor{ForestGreen}{\textbf{$\downarrow$ 9.0\%}}              \\ \bottomrule
\end{tabular}
}
\label{tab:pro_multilingual}
\vspace*{-1.5\baselineskip}
\end{table}

\subsection{Replaying Environment before Steps}\label{sec:return}
Once a target step is selected, \ours must restore the replay context and environment state \textit{before} this step, as shown in Figure \ref{fig:overview}. Since prior trajectories are recorded in the archive, the replay context can be retrieved directly. For the environment state, storing full snapshots is storage-intensive, and replaying entire action sequences can be inefficient, as some actions (\eg running test suites) are costly. To address this, we observe that software agents modify the environment mostly through text edits and file creation/deletion in the repository. Consequently, we start by checking with pattern matching whether agents' actions have mutated non-repo state (\eg package installs): if unchanged, we will only record the repository diff for each step: when resuming from a step, the environment is restored by applying the stored diffs to the repository, avoiding action replay and improving efficiency; otherwise, \ours will replay the original action sequence to restore the environment.

\subsection{Branching at Steps}\label{sec:explore}
After restoration, the agent resumes exploration by sampling a new step to \textit{replace the selected step} and then continuing to complete a new trajectory, as shown in Figure \ref{fig:overview}. The resulting new trajectory, which combines the replay context and generated suffix, is added to the archive for future reuse.

\textbf{Overhead analysis.}
Unlike existing works~\cite{antoniades2024swe,zeng2025satorisweevolutionarytesttimescaling} which require costly tree search or model training, \ours introduces negligible overhead. Firstly, step selection only requires lightweight system operations (\eg loading trajectory files) and simple numeric computations, which do not introduce any meaningful overhead. Likewise, the overhead of returning to the steps is negligible: it only requires applying a diff file, which takes on the order of seconds and is insignificant compared to the cost of generating full trajectories.

\begin{table*}[]
\caption{Ablation on \ours designs. When selection strategies are disabled ({\color{Red} \xmark}), the method defaults to uniform random selection.}
\vspace*{-0.3\baselineskip}
\centering
\resizebox{1.9\columnwidth}{!}{
\begin{tabular}{@{}cccccccc@{}}
\toprule
\multirow{2}{*}{\begin{tabular}[c]{@{}c@{}}Naive\\Scaling\end{tabular}} & \multirow{2}{*}{\begin{tabular}[c]{@{}c@{}}Trajectory\\Filtering\end{tabular}} & \multirow{2}{*}{\begin{tabular}[c]{@{}c@{}}Selection \#1:\\Representation\end{tabular}} & \multirow{2}{*}{\begin{tabular}[c]{@{}c@{}}Selection \#2:\\Reasoning\end{tabular}} & \multicolumn{1}{c}{Performance} & \multicolumn{3}{c}{Efficiency}         \\ \cmidrule(l){5-8} 
                               &                                       &                                                       &                                                   & \multicolumn{1}{c}{\% Resolved $\uparrow$} & \multicolumn{1}{c}{\# Input (k) $\downarrow$} & \multicolumn{1}{c}{\# Output (k) $\downarrow$} & \multicolumn{1}{c}{Avg. Cost (\$) $\downarrow$} \\ \midrule
{\color{Green} \CheckmarkBold}       &    \textbf{--}       &        \textbf{--}     &       \textbf{--}      & 52.0\%                          & 16814.7                                 & 149.4                                    & 1.73                               \\ \midrule
\textbf{--}        &         {\color{Red} \xmark}          &      {\color{Red} \xmark}    &          {\color{Red} \xmark}        & 56.0\%                          & 17373.3                                 & 132.1                                    & 1.78                               \\
\addlinespace[0.1em]
\cdashline{1-8}
\addlinespace[0.3em]
\textbf{--}       &       {\color{Green} \CheckmarkBold}      &                   {\color{Red} \xmark}         &       {\color{Red} \xmark}               & 56.0\%                          & 14952.7                                 & \textbf{117.4}                           & 1.53                               \\
\addlinespace[0.1em]
\cdashline{1-8}
\addlinespace[0.3em]
\textbf{--}          &       {\color{Green} \CheckmarkBold}        &          {\color{Green} \CheckmarkBold}                &        {\color{Red} \xmark}           & 58.0\%                          & 15083.4                                 & 126.7                                    & 1.55                               \\ \midrule
\textbf{--}           &       {\color{Green} \CheckmarkBold}         &        {\color{Green} \CheckmarkBold}                     &    {\color{Green} \CheckmarkBold}                            & $\,$\textbf{60.0\%}                 & \textbf{14795.3}                        & 131.8                                 & \textbf{1.52}                      \\ \bottomrule
\end{tabular}
}
\label{tab:ablation_main}
\vspace*{-\baselineskip}
\end{table*}

\section{Evaluation}\label{sec:evaluation}

\subsection{Experimental Setup}\label{sec:setup}

\textbf{Implementation.}
\ours is generalizable across various agent scaffolds. In our experiments, we implement \ours on top of two popular frameworks: mini-SWE-agent~\cite{yang2024swe} and Live-SWE-agent~\cite{xia2025live}. We evaluate performance using two closed-source and one open-source LLM backends: Gemini-2.5-Pro~\cite{gemini25pro}, Gemini-3-Pro\footnote{https://deepmind.google/models/gemini/}, and Devstral-Small-2\footnote{https://mistral.ai/news/devstral-2-vibe-cli}. We sample ten patches per issue and employ an Agentless-style selection mechanism to submit the final solution~\cite{xia2024agentless}. More details are provided in Appendix \ref{sec:appendix_hyperparameter}.

\textbf{Datasets.}
We evaluate \ours on three primary datasets: SWE-Bench Verified, SWE-Bench Pro~\cite{deng2025swe}, and SWE-Bench Multilingual. Following prior works~\cite{zhang2025darwingodelmachineopenended, xia2025live}, we utilize SWE-Bench Verified mini\footnote{\scriptsize{https://github.com/mariushobbhahn/SWEBench-verified-mini}}, a 50-problem subset of SWE-Bench Verified, for our ablation studies in \S\ref{sec:ablation}. Further dataset details are presented in Appendix \ref{sec:appendix_benchmark}.

\textbf{Baselines and Metrics.}
Our primary baseline is \textit{naive scaling}, where all ten trajectories are sampled from scratch. In contrast, \ours generates the initial trajectory from scratch and performs exploration-exploitation sampling (\S\ref{sec:select_adaptive}) for the remaining nine. We do not compare directly with SWE-Search or Satori-SWE as they rely on pre-defined tool templates and are incompatible with the open-ended action spaces of modern agents. Instead, we consider an LLM-as-a-Judge baseline (Table \ref{tab:ablation_reward}) where we replace step selection in \ours with LLM-as-a-Judge to estimate step quality and select steps based on the score.

Following standard practices~\cite{xia2024agentless,xia2025live}, we report two metrics: (1) \textbf{Performance}, measured by the resolve rate; and (2) \textbf{Efficiency}, measured by the average number of input/output tokens and the cost per problem. Costs account for \textit{explicit} prompt caching, as \ours maximizes prefix sharing. See Appendix \ref{sec:appendix_metric} for cost modeling details.

\begin{table}[t]
\caption{Ablation on different step selection methods.}
\vspace*{-0.3\baselineskip}
\centering
\resizebox{0.9\columnwidth}{!}{
\begin{tabular}{@{}ccccc@{}}
\toprule
\multirow{2}{*}{Step Selection}       & \multirow{2}{*}{\% Resolved $\uparrow$}       & \multicolumn{3}{c}{Avg. Cost (\$) $\downarrow$} \\ \cmidrule(l){3-5}
       &        &  Agent  &  Reward & Total \\ \midrule
Random               & 56.0\%          & 1.53       & 0 & 1.53 \\
LLM-as-a-Judge       & 54.0\%          &  1.44        &  1.87  & 3.31 \\ \midrule
\ours & $\,$\textbf{60.0\%} & 1.52 & 0 & \textbf{1.52} \\ \bottomrule
\end{tabular}
}
\label{tab:ablation_reward}
\vspace*{-1.5\baselineskip}
\end{table}

\subsection{Main Results}\label{sec:main}

\textbf{SWE-Bench Verified.}
Table \ref{tab:main} presents the comparative results of \ours and naive test-time scaling on SWE-Bench Verified. We observe that \ours consistently achieves a higher resolve rate while incurring lower average costs across all agent scaffolds and LLM backends. Specifically, \ours reduces the computational cost of naive scaling by up to 17.4\% while improving performance by up to 3.8\%. These results highlight the ability of \ours to deliver consistent efficiency gains and performance improvements without requiring specialized reward models.

\textbf{SWE-Bench Pro and Multilingual.}
We conduct additional evaluations on the more complex SWE-Bench Pro and Multilingual benchmarks, using Devstral-Small-2 with mini-SWE-agent. Table \ref{tab:pro_multilingual} shows that \ours maintains its superiority over naive scaling, achieving an up to 22.6\% performance gain with an up to 9.0\% reduction in cost. These results confirm that \ours generalizes effectively to enterprise-level multilingual SWE problems.

\subsection{Ablation Experiments}\label{sec:ablation}

\textbf{Ablation of \ours Components.}
We perform ablation studies to validate the design choices of \ours, focusing on three core components: trajectory filtering, step grouping via abstraction, and step selection based on reasoning intensity. Following the setup in \S\ref{sec:setup}, all ablations use the Devstral-Small-2 backend with Live-SWE-agent as it achieves a better performance on SWE-Bench Verified. As shown in Table \ref{tab:ablation_main}, the full \ours implementation achieves the best results, with a 60.0\% resolve rate and an average cost of \$1.52 per problem. Progressively disabling these components leads to degradation in both performance and efficiency. These findings underscore the necessity of each component in the \ours pipeline.

\begin{figure}[t]
\centering
\includegraphics[width=0.85\linewidth]{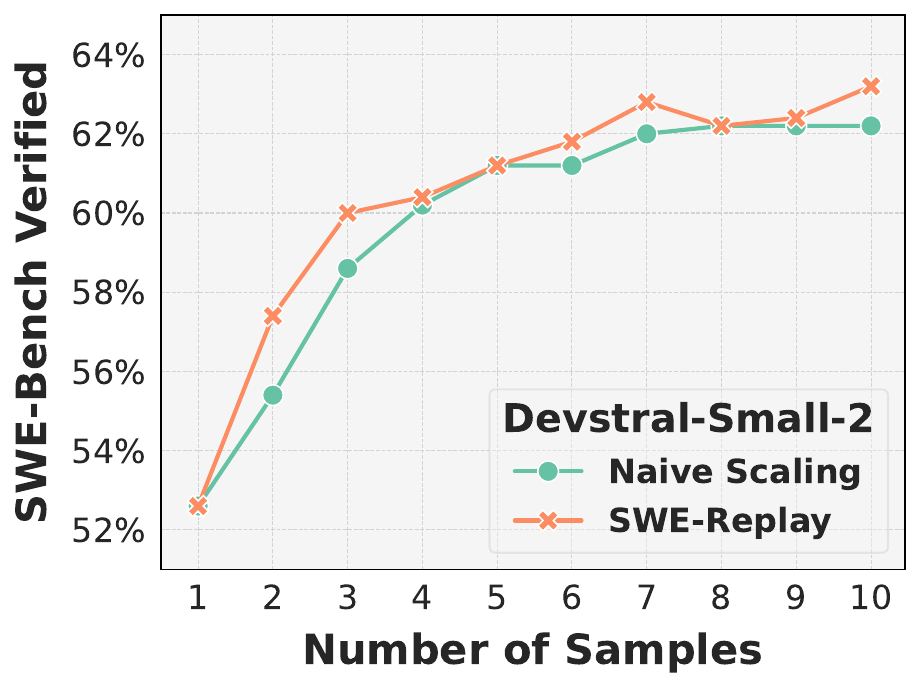}
\vspace*{-0.3\baselineskip}
\caption{Scaling curve of \ours for Devstral-Small-2.}
\label{fig:scaling}
\vspace*{-1.5\baselineskip}
\end{figure}

\textbf{Comparison of LLM-as-a-Judge and \ours.}
We experiment to study whether we can replace the step selection mechanism of \ours with LLM-as-a-Judge. Specifically, we explore directly using LLM-as-a-Judge to estimate the quality of each step in the archive and then sampling one based on the LLM-generated quality score. More details are included in Appendix~\ref{sec:appendix_reward}. As shown in Table \ref{tab:ablation_reward}, LLM-as-a-Judge performs worse than \ours while requiring additional costs to estimate the quality score using an LLM, demonstrating that step selection in \ours is crucial to boost the scaling performance without being impacted by potentially inaccurate LLM-as-a-Judge.

\begin{figure*}[t]
\centering
\includegraphics[width=0.9\linewidth]{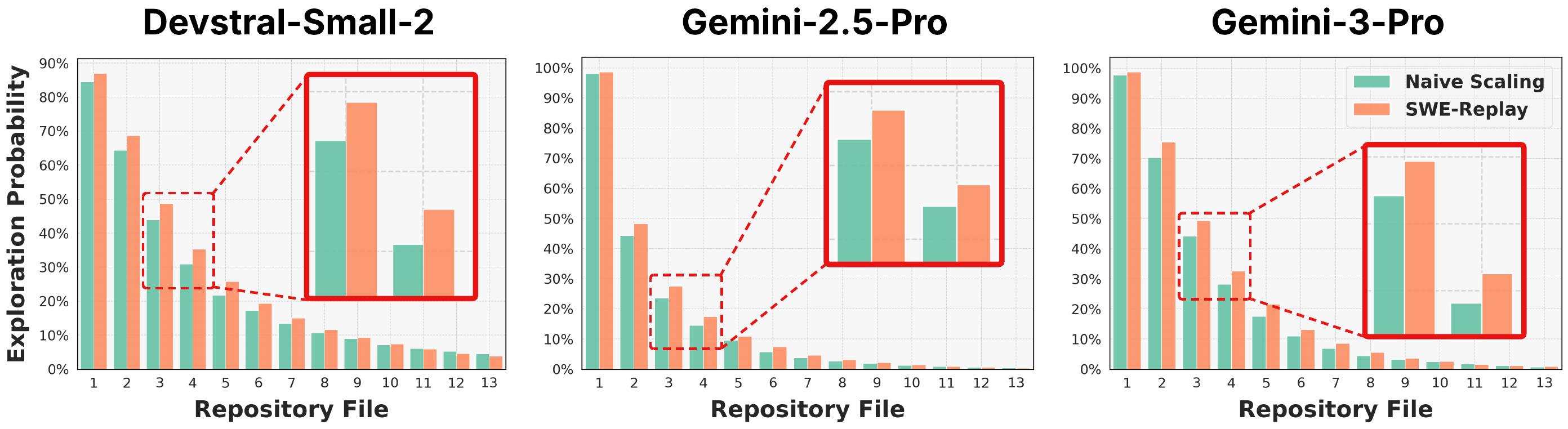}
\caption{Distribution of repository files explored by \ours and naive scaling.}
\label{fig:distribution}
\vspace*{-\baselineskip}
\end{figure*}

\textbf{Scaling analysis of \ours.}
We evaluate how well the resolve rate on SWE-Bench Verified scales with the number of samples in \ours. Figure \ref{fig:scaling} shows that increasing the number of samples enhances its performance steadily, while outperforming naive scaling consistently. 

More ablations on the impact of different step representations (\S\ref{sec:ablation_abstraction}), reasoning intensity representations (\S\ref{sec:ablation_reasoning}), and normalizations (\S\ref{sec:ablation_normalization}) are detailed in the Appendix.

\section{Discussion}\label{sec:discussion}

\begin{figure}[t]
\centering
\includegraphics[width=0.91\linewidth]{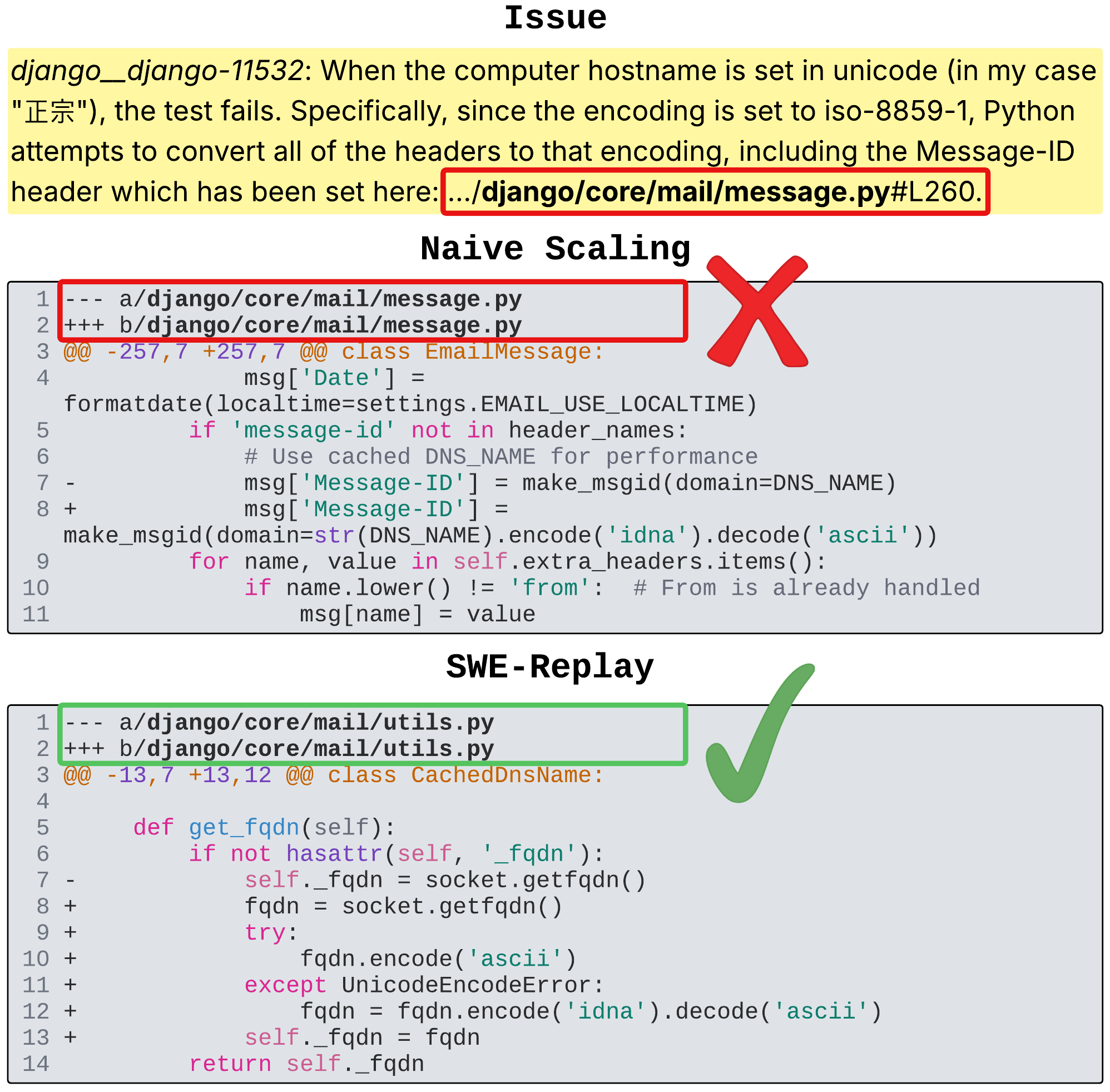}
\vspace*{-0.3\baselineskip}
\caption{Example of fixing failure due to inadequate exploration. Naive scaling fixated on \texttt{message.py} (incorrect), whereas \ours successfully branched out to explore \texttt{utils.py} (correct).}
\label{fig:exploration_example}
\vspace*{-\baselineskip}
\end{figure}

\subsection{\ours Diversifies Repository Exploration}
A core intuition motivating the step selection mechanism of \ours (\S\ref{sec:select_abstract}) is to explore repository spaces that are largely ignored in prior trials. To verify whether \ours can diversify the search space, we analyze the distribution of repository files visited by the agent on SWE-Bench Verified. Specifically, for each problem, we extract the set of files explored during the scaling process. To aggregate distributions across problems with varying file structures, we normalize file identities by ranking them based on visitation frequency. For a given problem, the most frequently visited file is assigned rank 1, the second most visited rank 2, and so on. This normalization allows us to average the exploration frequency of files across different problems. Thus, effectively accessing the "tail" of the distribution indicates greater diversity.

Figure~\ref{fig:distribution} illustrates the distribution of repository files across different LLMs. A key observation is that while naive scaling and \ours exhibit similar probabilities of exploring the most frequent (top-ranked) files, \ours consistently increases the probability of visiting less frequent files (the long tail). These results confirm that \ours successfully encourages the agent to investigate less frequently visited files. The case study in Figure~\ref{fig:exploration_example} demonstrates the practical advantage of this diversity. In this instance, the issue description explicitly points to \texttt{message.py} as the potential location of the bug. Consequently, naive scaling fixates on this file, generating a patch that fails to address the root cause. In contrast, \ours breaks away from this local optimum to explore \texttt{utils.py}, successfully fixing the issue at a lower level. This highlights the critical importance of exploring a diverse repository space to locate the correct patch.

\subsection{Theoretical Intuition of \ours}
\label{sec:theoretical_analysis}

We present the theoretical intuition of \ours by considering a simplified scenario to model the efficiency of \ours compared to naive scaling. We assume a challenging SWE task characterized by the following properties:

\vspace*{-0.8\baselineskip}

\begin{enumerate}[leftmargin=1em]
    \setlength{\parskip}{0pt}
    \item All sampled trajectories consist of exactly $N$ steps.
    \item There exists a single optimal path (the "correct trajectory") that solves the issue.
    \item \textbf{Recoverability:} If the archive contains the correct trajectory and \ours branches from any step within it, the agent is guaranteed to fix the issue.
    \item \textbf{Hardness:} The probability $p$ of generating this correct trajectory from scratch in a single trial is small ($p \ll 1$).
\end{enumerate}

\vspace*{-0.8\baselineskip}

Let us define the probabilities for the $t$-th trial ($t \geq 2$):

\vspace*{-0.8\baselineskip}

\begin{enumerate}[leftmargin=1em]
    \setlength{\parskip}{0pt}
    \item $P_{\text{replay}}^{(t)}$: The probability of \ours generating the correct solution at trial $t$.
    \item $P_{\text{select}}^{(t)}$: The probability that the step selection mechanism chooses a step belonging to the correct trajectory, if the correct trajectory already exists in the archive.
    \item $P'_{\text{select}}(t)$: The probability of selecting a step from an incorrect trajectory but still randomly stumbling upon the solution (negligible for hard tasks).
\end{enumerate}

\vspace*{-0.8\baselineskip}

We aim to show: \textbf{provided $P_{\text{select}}^{(t)}$ exceeds the probability of random selection, \ours satisfies $P_{\text{replay}}^{(t)} \geq p$}.

At each trial, \ours generates from scratch with probability 0.5 and also replays from archive with probability 0.5. The marginal probability of success at trial $t$ is:
\vspace*{-0.3\baselineskip}
\begin{equation}
\begin{aligned}
&P_{\text{replay}}^{(t)} = \frac{1}{2}p + \frac{1}{2} \Big[ \underbrace{\mathbb{P}(\text{Archive has correct})}_{= 1-(1-p)^{t-1}} \cdot P_{\text{select}}^{(t)} \\
&\quad + \Big[ \underbrace{\mathbb{P}(\text{Archive has only incorrect})}_{= (1-p)^{t-1}} \cdot P'_{\text{select}}(t) \Big].
\end{aligned}
\end{equation}
\vspace*{-\baselineskip}

Dropping the negligible term $P'_{\text{select}}(t)$ and requiring $P_{\text{replay}}^{(t)} \geq p$, we obtain the inequality:
\vspace*{-0.3\baselineskip}
\begin{equation}
\frac{1}{2}p + \frac{1}{2} \left(1-(1-p)^{t-1}\right) P_{\text{select}}^{(t)} \geq p.
\end{equation}
\vspace*{-\baselineskip}

Rearranging for $P_{\text{select}}^{(t)}$:
\vspace*{-0.3\baselineskip}
\begin{equation}
P_{\text{select}}^{(t)} \geq \frac{p}{1-(1-p)^{t-1}}.
\end{equation}
\vspace*{-1.5\baselineskip}

Since $p$ is small (Assumption 4), we can approximate the denominator using the first-order Taylor expansion, $(1-p)^{t-1} \approx 1 - (t-1)p$. This simplifies the condition to:
\vspace*{-0.3\baselineskip}
\begin{equation}
\label{eq:condition}
P_{\text{select}}^{(t)} \gtrsim \frac{p}{(t-1)p} = \frac{1}{t-1}.
\end{equation}
\vspace*{-\baselineskip}

This lower bound, $\frac{1}{t-1}$, corresponds precisely to the probability of \textit{random selection}. At trial $t$, the archive contains $t-1$ trajectories, and the probability of uniformly selecting a step from the correct trajectory is $\frac{N}{(t-1)N} = \frac{1}{t-1}$. Consequently, inequality (\ref{eq:condition}) implies that as long as the archive has contained the correct trajectory and the \ours selection mechanism assigns higher probability to steps on the correct trajectory than uniform random sampling, \ours will outperform naive scaling ($P_{\text{replay}}^{(t)} \geq p$). In practice, by filtering low-quality trajectories and prioritizing reasoning-intensive steps, \ours biases selection toward the correct trajectory, as validated in \S\ref{sec:ablation}.
\section{Related Work}\label{sec:related}

\subsection{Scaffolding for Software Engineering Agents}
Drawing inspiration from human debugging loops, where developers learn from environmental feedback such as test failures, early interactive solutions~\cite{xia2024automated, chen2023teaching} pioneered bug-fixing via Large Language Models (LLMs). Since those initial efforts, a significant body of research regarding bug fixing and general coding tasks has focused on automatically providing LLMs with greater context through multi-turn conversations~\cite{yang2024swe,wang2024openhands}. More recently, foundation LLMs have demonstrated substantial improvements in reasoning and tool usage, enabling the development of feedback-driven solutions that leverage these emergent capabilities~\cite{carbonneaux2025cwm,ding2025empoweringmultiturntoolintegratedreasoning}.
In March 2024, the field saw the release of Devin AI\footnote{https://cognition.ai/blog/introducing-devin}, the first software engineer agent designed to autonomously complete end-to-end software tasks, such as resolving GitHub issues. The initial release of Devin demonstrated impressive performance on the SWE-Bench dataset, which consists of thousands of real-world GitHub issues~\cite{jimenez2023swe}. Following this milestone, numerous dedicated software agent scaffolds have been proposed, including SWE-agent~\cite{yang2024swe}, OpenHands~\cite{wang2024openhands}, AutoCodeRover~\cite{zhang2024autocoderover}, Trae Agent~\cite{gao2025trae}, and Live-SWE-agent~\cite{xia2025live}. These systems typically equip LLMs with a suite of coding tools, enabling the models to determine the necessary actions to complete real-world software tasks autonomously.

\subsection{Test-Time Scaling for Software Engineering Agents}
Allocating test-time computation to sample multiple candidate patches has emerged as a highly effective strategy for software engineering agents. Existing works demonstrate that repeated sampling with smaller LLMs often yields superior solution coverage compared to single-shot outputs from larger models~\cite{brown2024large}. This approach is particularly advantageous in software engineering tasks, where execution-based feedback facilitates the rigorous selection of valid code~\cite{ehrlich2025codemonkeys,ma2025thinkinglongerlargerenhancing}. To mitigate the computational costs associated with such scaling, SWE-Search~\cite{antoniades2024swe} has integrated tree search algorithms to prune unpromising trajectories early, relying on a separate agent to estimate quality. Similarly, Satori-SWE~\cite{zeng2025satorisweevolutionarytesttimescaling} proposes to achieve sample-efficient test-time scaling by using LLM to self-improve the scores of its prior generations assigned by a reward model. However, this work presents two significant drawbacks: (1) quality scores are often skewed by model overconfidence and miscalibration~\cite{son2024llm}, introducing noise into the scaling process; (2) SWE-Search and Satori-SWE are designed specifically for the tool designs of pipeline-based scaffolds (\eg Moatless) and not directly generalizable to modern agents. In contrast, \ours bypasses reliance on potentially inaccurate predictions, is naturally generalizable to all agentic scaffolds, and employs a streamlined select-and-replay mechanism to ensure scalability.
\section{Conclusion}\label{sec:conclusion}
In this work, we presented \ours, a novel efficient test-time scaling framework that reconciles the tension between computational efficiency and solution quality for software engineering agents by recycling previously sampled trajectories, dynamically choosing to either explore from scratch or exploit archived experience by branching at critical intermediate steps. Critical intermediate steps are selected based on the potential and reasoning significance of repository exploration, rather than LLM-as-a-Judge.

Through evaluation on SWE-Bench Verified, SWE-Bench Pro, and SWE-Bench Multilingual across multiple LLM backends and agentic scaffolds, we demonstrated that: (1) \ours reduces computational costs by up to 17.4\% while improving resolve rates by up to 3.8\% on SWE-Bench Verified, and (2) \ours showcases consistent generalizability towards diverse types of problems in SWE-Bench Pro and SWE-Bench Multilingual. We further validate empirically that, compared with naive scaling, \ours enables agents to explore more diverse, less visited repository spaces. These empirical gains are supported by our theoretical intuition, which provides a foundational understanding of how replaying optimizes the quality of exploration.

\section*{Impact Statement}

This paper presents work whose goal is to advance the field of Machine Learning. There are many potential societal consequences of our work, none which we feel must be specifically highlighted here.

\bibliography{example_paper}
\bibliographystyle{icml2026}

\newpage
\appendix
\onecolumn
\section{Appendix}\label{sec:appendix}

\begin{algorithm}[h]
\caption{\ours}
\label{alg:swe_replay}
\begin{algorithmic}[1]
\REQUIRE Issue Description $D$, Agent $\mathcal{A}$, Budget $N$
\ENSURE Final Patch $P^*$

\STATE Initialize Trajectory Archive $\mathcal{T} \leftarrow \emptyset$

\FOR{$i = 1$ to $N$}
    \STATE $mode \leftarrow \text{EXPLORE}$
    
    \IF{$i > 1$}
        \STATE $mode \leftarrow \text{Bernoulli}(0.5) \ ? \ \text{EXPLORE} : \text{EXPLOIT}$
    \ENDIF

    \IF{$mode = \text{EXPLORE}$}
        \STATE $S_{start} \leftarrow \text{InitializeEnv}(D)$
        \STATE $\tau_{new} \leftarrow \mathcal{A}.\text{Run}(S_{start}, \text{context}=\emptyset)$
    \ELSE
        \STATE $s_{selected} \leftarrow \text{SelectStep}(\mathcal{T})$ 
        \STATE $S_{resumed} \leftarrow \text{RestoreEnv}(s_{selected})$ 
        \STATE $C_{resumed} \leftarrow \text{GetContext}(s_{selected})$
        \STATE $\tau_{suffix} \leftarrow \mathcal{A}.\text{Run}(S_{resumed}, C_{resumed})$
        \STATE $\tau_{new} \leftarrow \text{Concatenate}(C_{resumed}, \tau_{suffix})$
    \ENDIF

    \STATE $\mathcal{T} \leftarrow \mathcal{T} \cup \{ \tau_{new} \}$
\ENDFOR

\STATE $\mathcal{P}_{candidates} \leftarrow \{ \text{GetPatch}(\tau) \mid \tau \in \mathcal{T} \}$
\STATE $\mathcal{P}_{valid} \leftarrow \text{FilterTestFailures}(\mathcal{P}_{candidates})$
\STATE $P^* \leftarrow \text{MajorityVote}(\mathcal{P}_{valid})$
\end{algorithmic}
\end{algorithm}

\subsection{Implementation Details}\label{sec:appendix_hyperparameter}
Following the default setting of mini-SWE-agent~\cite{yang2024swe}, we set a maximum step limit of 250 for Gemini-2.5-Pro and Gemini-3-Pro. For Devstral-Small-2, we set a maximum step limit of 128 to accelerate the experiments. We ignore the cost limit in our experiments to ensure that the agents have enough opportunities to explore the repositories without being affected by the external model pricing. Based on our experience, we set a temperature of 0.8 for Gemini-2.5-Pro and a temperature of 0.2 for Devstral-Small-2 and Gemini-3-Pro to achieve the best performance. This also gives us a chance to observe the performance of \ours under different levels of temperatures. We sample ten trajectories per problem during scaling for all the experiments on SWE-Bench Verified and SWE-Bench Multilingual. Due to the large size of SWE-Bench Pro, we sample five trajectories per problem on this benchmark.

While we follow the Agentless~\cite{xia2024agentless} pipeline to select the final patch, we remove LLM-generated reproduction tests in the pipeline to remove any potential noise. So our final selection pipeline is: (1) run the regression tests for all the candidate patches, (2) filter out all the candidate patches with regression failures, and (3) select the final patch out of the remaining ones using majority voting.

\subsection{Benchmark Details}\label{sec:appendix_benchmark}
The SWE-Bench Verified benchmark comprises 500 SWE tasks, where the objective is to modify a repository based on a provided description. These instances are human-validated to ensure that the problem descriptions contain sufficient information for a solution. To further test the agent on realistic, enterprise-level complexity, we also employ SWE-Bench Pro and SWE-Bench Multilingual. SWE-Bench Pro comprises 731 publicly available problems and differs from SWE-Bench Verified by presenting higher difficulty levels across a diverse range of repositories and programming languages. Similarly, SWE-Bench Multilingual is a benchmark of 300 software engineering tasks across 9 programming languages (JavaScript, TypeScript, Rust, Ruby, Go, C/C++, PHP, and Java).

SWE-Bench Verified mini is a subset of SWE-Bench Verified that uses 50 instead of 500 data points, requires 5GB instead of 130GB of storage, and has approximately the same distribution of performance, test pass rate, and difficulty as the original dataset. It includes the following problems from SWE-Bench Verified:

\begin{multicols}{3}
\begin{itemize}
\item \texttt{django\_\_django-11790}
\item \texttt{django\_\_django-11815}
\item \texttt{django\_\_django-11848}
\item \texttt{django\_\_django-11880}
\item \texttt{django\_\_django-11885}
\item \texttt{django\_\_django-11951}
\item \texttt{django\_\_django-11964}
\item \texttt{django\_\_django-11999}
\item \texttt{django\_\_django-12039}
\item \texttt{django\_\_django-12050}
\item \texttt{django\_\_django-12143}
\item \texttt{django\_\_django-12155}
\item \texttt{django\_\_django-12193}
\item \texttt{django\_\_django-12209}
\item \texttt{django\_\_django-12262}
\item \texttt{django\_\_django-12273}
\item \texttt{django\_\_django-12276}
\item \texttt{django\_\_django-12304}
\item \texttt{django\_\_django-12308}
\item \texttt{django\_\_django-12325}
\item \texttt{django\_\_django-12406}
\item \texttt{django\_\_django-12708}
\item \texttt{django\_\_django-12713}
\item \texttt{django\_\_django-12774}
\item \texttt{django\_\_django-9296}
\item \texttt{sphinx-doc\_\_sphinx-10323}
\item \texttt{sphinx-doc\_\_sphinx-10435}
\item \texttt{sphinx-doc\_\_sphinx-10466}
\item \texttt{sphinx-doc\_\_sphinx-10673}
\item \texttt{sphinx-doc\_\_sphinx-11510}
\item \texttt{sphinx-doc\_\_sphinx-7590}
\item \texttt{sphinx-doc\_\_sphinx-7748}
\item \texttt{sphinx-doc\_\_sphinx-7757}
\item \texttt{sphinx-doc\_\_sphinx-7985}
\item \texttt{sphinx-doc\_\_sphinx-8035}
\item \texttt{sphinx-doc\_\_sphinx-8056}
\item \texttt{sphinx-doc\_\_sphinx-8265}
\item \texttt{sphinx-doc\_\_sphinx-8269}
\item \texttt{sphinx-doc\_\_sphinx-8475}
\item \texttt{sphinx-doc\_\_sphinx-8548}
\item \texttt{sphinx-doc\_\_sphinx-8551}
\item \texttt{sphinx-doc\_\_sphinx-8638}
\item \texttt{sphinx-doc\_\_sphinx-8721}
\item \texttt{sphinx-doc\_\_sphinx-9229}
\item \texttt{sphinx-doc\_\_sphinx-9230}
\item \texttt{sphinx-doc\_\_sphinx-9281}
\item \texttt{sphinx-doc\_\_sphinx-9320}
\item \texttt{sphinx-doc\_\_sphinx-9367}
\item \texttt{sphinx-doc\_\_sphinx-9461}
\item \texttt{sphinx-doc\_\_sphinx-9698}
\end{itemize}
\end{multicols}

\subsection{Metric Details}\label{sec:appendix_metric}
To evaluate computational efficiency, we record the cumulative number of input and output tokens across all ten trajectories for each problem. However, raw token counts fail to capture the practical benefits of \textit{prompt caching}, an inference optimization that stores the processed states of repeated text sequences to reduce latency and cost.

The branching mechanism of \ours, which resumes execution from intermediate steps, naturally yields substantial shared prefixes, thereby maximizing the utility of prompt caching. Although current commercial LLM backends (e.g., Gemini-2.5-Pro, Gemini-3-Pro) support caching, their default \textit{implicit} prompt caching policies do not always guarantee deterministic cost savings. Consequently, to provide a precise assessment of the cost reductions enabled by our method, we report the cost after enabling \textit{explicit} prompt caching. Specifically, we store all the prior generations in the prompt cache during scaling and report the cost of input/output tokens from LLMs.

\subsection{Ablation Details for LLM-as-a-Judge Variant}\label{sec:appendix_reward}
To replace the original step selection mechanism with LLM-as-a-Judge in \ours, we use the same LLM that the agent is using (\ie Devstral-Small-2 in this experiment) to generate a quality score $q_i\in[0,1]$ for each step $s_i$ in the archive. Then, similar to \S\ref{sec:select_reasoning}, we assign sampling probabilities for each step as follows:
\begin{equation}
p(s_i) = \mathrm{softmax}\!\left(q_i\right)
= \frac{e^{q_i}}{\sum_{j} e^{q_j}}.
\label{eq:select_step}
\end{equation}

In our experiment, we use Devstral-Small-2 to estimate a quality score for each step based on the following prompt:

\begin{promptbox}{Quality estimate prompt for LLM-as-a-Judge}
\begin{Verbatim}[breaklines=true]
Your task is to evaluate the quality of a middle step generated by an AI agent in a long trajectory. Your evaluation will focus on whether the step was well-constructed, whether the resulting execution outcome is relevant and useful for solving the problem at hand, and whether the analysis in the step is appropriate and helpful.

Here is the full trajectory:
"""
{trajectory}
"""

Here is the step you need to evaluate:
"""
{step}
"""

On the scale of 0 to 1 with a scale of 0.01, where 0 is extremely bad and 1 is extremely good, rate the quality of this step in the format of "score: ".
\end{Verbatim}
\end{promptbox}
\captionsetup{type=figure}
\captionof{figure}{
Prompt for Devstral-Small-2 to estimate the quality score for each step in the mini-SWE-agent trajectories.
}
\label{fig:reward_prompt}

\subsection{Effectiveness of File-Level Abstraction}\label{sec:ablation_abstraction}

We evaluate different levels of granularity for the abstract representation of steps (\S\ref{sec:select_abstract}): (1) \textbf{Line}: the set of all raw observations before the step; (2) \textbf{Method}: the set of all explored methods before the step; and (3) \textbf{File}: the set of all explored repository files before the step. As shown in Table \ref{tab:ablation_abstraction}, the file-level abstraction yields the highest resolve rate (60.0\%) and the lowest cost (\$1.52). In contrast, both method-level and line-level abstraction perform worse than file-level. This suggests that line-level and method-level representations are too granular and sparse, leading to fragmented groups that hinder effective reuse. File-level abstraction offers the optimal balance, grouping semantically related steps to maximize the benefits of exploration-exploitation.

\begin{table}[h]
\caption{Ablation on different step representations.}
\centering
\begin{tabular}{@{}ccc@{}}
\toprule
Abstraction Level & \% Resolved $\uparrow$    & Avg. Cost (\$) $\downarrow$ \\ \midrule
Line                                                           &      58.0\%           &  1.58              \\
Method                                                         &      $\,$\textbf{60.0\%}       &       1.57         \\ \midrule
File                                                           & $\,$\textbf{60.0\%} & \textbf{1.52}  \\ \bottomrule
\end{tabular}
\label{tab:ablation_abstraction}
\vspace*{-\baselineskip}
\end{table}

\subsection{Impact of Reasoning Intensity Representation}\label{sec:ablation_reasoning}

We verify our design choice for representing reasoning intensity (\S\ref{sec:select_reasoning}) by comparing two variants: using the raw length of reasoning content and our proposed method, which uses the number of reasoning paragraphs. As shown in Table \ref{tab:ablation_reasoning}, using the number of paragraphs yields the largest performance gains (60.0\%) while minimizing cost (\$1.52). This suggests that raw length is a noisy proxy for reasoning quality. In contrast, paragraph count serves as a lightweight yet accurate heuristic for identifying reasoning-intensive steps that are worthy of replay.

\begin{table}[h]
\caption{Ablation on different reasoning intensity representations.}
\centering
\begin{tabular}{@{}ccc@{}}
\toprule
Reasoning Intensity & \multicolumn{1}{c}{\% Resolved $\uparrow$} & \multicolumn{1}{c}{Avg. Cost (\$) $\downarrow$} \\ \midrule
Length                                                  & 58.0\%                          & 1.55                               \\ \midrule
\# Paragraphs                                                   & $\,$\textbf{60.0\%}                 & \textbf{1.52}                      \\ \bottomrule
\end{tabular}
\label{tab:ablation_reasoning}
\vspace*{-0.5\baselineskip}
\end{table}

\subsection{Impact of Softmax Normalization}\label{sec:ablation_normalization}

\ours utilizes a normalized probability distribution to balance exploration and exploitation during step selection. We compare two normalization techniques: \textit{Unit Sum} ($\frac{v_i}{\sum v_i}$) and \textit{Softmax} ($\frac{e^{v_i}}{\sum e^{v_i}}$). In the context of our step selection mechanism, Unit Sum tends to produce a sharper distribution that may prematurely converge on specific steps. In contrast, Softmax produces a flatter distribution that better preserves diversity. As shown in Table \ref{tab:ablation_normalization}, replacing Softmax with Unit Sum results in a marked reduction in both performance (56.0\% vs. 60.0\%) and efficiency (\$1.63 vs. \$1.52), confirming that Softmax normalization is critical for maintaining an effective diversity of sampling candidates.

\begin{table}[h]
\caption{Ablation on normalization methods for step selection.}
\centering
\begin{tabular}{@{}ccc@{}}
\toprule
Normalization & \multicolumn{1}{c}{\% Resolved $\uparrow$} & \multicolumn{1}{c}{Avg. Cost (\$) $\downarrow$} \\ \midrule
Unit Sum                                                       & 56.0\%                          & 1.63                               \\ \midrule
Softmax                                                        & $\,$\textbf{60.0\%}                 & \textbf{1.52}                      \\ \bottomrule
\end{tabular}
\label{tab:ablation_normalization}
\vspace*{-\baselineskip}
\end{table}

\end{document}